\documentclass[aps,twocolumn,showkeys]{revtex4}  
\usepackage[]{graphicx}

\begin{document} 

\title{Surface EM waves on 1D Photonic Crystals}

\author{G. V. Morozov}
\affiliation{School of Mathematics, University of Bristol\\
Bristol, BS8 1TW United Kingdom}

\author{D. W. L. Sprung}
\affiliation{Department of Physics and Astronomy, McMaster University\\
Hamilton, Ontario L8S 4M1 Canada}

\author{J. Martorell}
\affiliation{Departament d'Estructura i Constituents de la Materia, Facultat
F\'{\i}sica, University of Barcelona\\  Barcelona 08028, Spain}

\date{June 27, 2006}

\begin{abstract}

We study surface states of 1D photonic crystals using a semiclassical 
coupled wave theory. Both TE and TM modes are treated. We derive 
analytic approximations that clarify the systematics of the 
dispersion relations, and the roles of the various parameters 
defining the crystal.

\end{abstract}


\maketitle

\section{INTRODUCTION}
Photonic crystals are artificial low-loss dielectric structures 
with periodic modulation of refractive index, which have attracted 
considerable attention in the last two decades. Due to Bragg reflection, 
electromagnetic (optical) waves cannot propagate through such structures 
in certain directions, at certain frequencies. Hence, photonic crystals 
can control the propagation of electromagnetic waves in novel ways, 
with obvious application to dielectric mirrors, dielectric waveguides, 
and dielectric laser cavities.

As a way to efficiently inject light into a photonic crystal (PC) 
waveguide, it has recently been proposed to use surface electromagnetic 
waves (SEW)\cite{Moreno2004,Kramper2004}. In those papers, the 
photonic crystal was a two dimensional array of rods, of infinite 
length normal to the plane of incidence. Instead, we have studied SEW 
on a semi-infinite one-dimensional (1D) photonic crystal sketched in 
Fig. 1.  While retaining all useful properties of 2D and 3D photonic 
crystals, a 1D dielectric periodic structure with high refractive 
index contrast is more attractive from a technological point of view. 

The usual theoretical methods for wave propagation in 1D photonic
crystals, including SEW , are the Floquet-Bloch modal formalism, coupled 
wave theory, and the transfer matrix method. Among these three, the 
coupled wave approach\cite{Kogelnik1969,Yariv1984,Yeh1987}
offers superior physical insight and gives simple analytical results 
in limiting cases. Unfortunately, the conventional coupled wave 
theory of Kogelnik fails in the case of high refractive index 
contrast, which is essential for a functional 1D photonic crystal. 

In this paper, we apply our recently developed semiclassical version 
of coupled wave theory\cite{Stolyarov1993,G1,G2} to SEW on 1D 
photonic crystals. The method is analytically almost as simple as 
conventional coupled wave theory, and is practically exact for the 
achievable ratios (e.g. 1.5:4.6) of the indices of refraction of the 
materials available to build devices. We present here a unified 
description of TE and TM SEW. A detailed account of the properties of 
the TE surface modes has recently been given by us in 
Ref.~\cite{Martorell2006}; here we complement these findings with 
those for TM modes, which are slightly more complex due to the presence of 
Brewster points in the bandgaps. As a result, we thoroughly clarify 
the systematics of solutions for surface EM waves in semi-infinite 1D 
photonic crystals. 

Our method is formally quite different from that recently 
presented in Ref.~\cite{Gaspar2004}, 
or those in Ref.~\cite{Petit1980}, so in Section II we provide a 
short summary of the transfer matrix approach, in the notation of  
our previous work\cite{G1}. In Section III we rederive the exact 
equations for SEW of TM modes and obtain from them various analytic 
approximations for a semi-infinite crystal. The analogous equations 
for TE modes were given in Ref.~\cite{Martorell2006}. With these in 
hand, we discuss systematics of SEW. In Section IV we apply the 
semiclassical approximations of Refs.~\cite{G1} and 
\cite{G2} to surface waves, and show that the second approximation 
is very accurate both for the dispersion relation and the bandgap 
boundaries.

\section{TRANSFER MATRIX METHOD FOR A PERIODIC CRYSTAL}

\begin{figure}[b]
\begin{center}
\begin{tabular}{c}
\includegraphics[height=7cm,width=8.5cm]{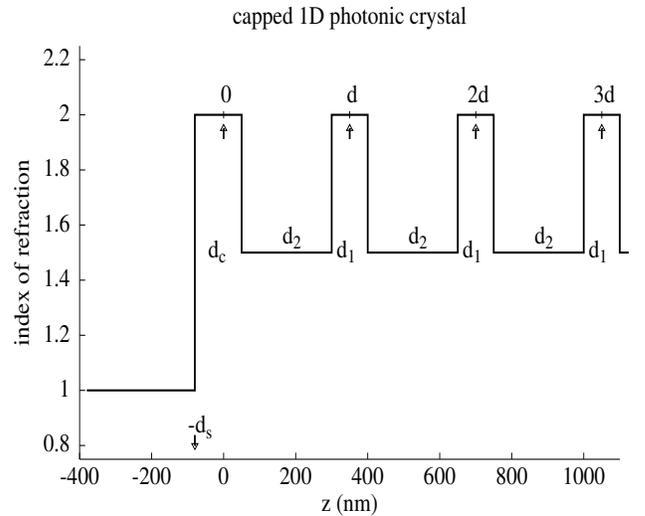}
\end{tabular}
\end{center}
\caption
{\label{dsfig01} 
Sketch of a typical 1D PC, for the case used in 
calculations ($n_1 = 2.0$, $d_1 = 100$ nm, $n_2 = 1.5$, $d_2 = 250$ nm,) 
A symmetric unit cell was chosen, and $d_s = 80$ nm.}
\end{figure}

We wish to describe surface states that form at the interface between 
a medium of low refractive index, $n_0$, and a semi-infinite 1-D 
photonic crystal with layers of refractive indices $n_1$ and $n_2$ 
and thicknesses $d_1$ and $d_2$. We choose a coordinate system in 
which the layers have normal vector along OZ. As shown in Fig. 
\ref{dsfig01}, the crystal is capped by a layer of the same material 
but different width, $d_c$. For convenience of presentation, we split 
this termination layer of index of refraction $n_1$ and width $d_c$ 
into two sublayers, of lengths $d_s + d_t = d_c$. The first sublayer 
extends from $z=-d_s$ to $z=0$. Then the periodic array that forms 
the 1D photonic crystal consists of ``cells'' each made of three 
uniform layers of widths $d_t$, $d_2$ and $d_1-d_t$ whose respective 
indices of refraction are $n_1$ , $n_2$ and $n_1$. (If $d_t = 
d_1/2$, the unit cell will have reflection symmetry, which simplifies 
some relations, but does not change any physical results.) The first 
cell, given index $0$,  ranges from $z=0$ to $z=d \equiv d_1+d_2$; 
the second is given index $1$, and ranges from $z=d$ to $2d$, etc.  
The p-th cell runs from $p d $ to $(p+1)d$ and has $n(z) = n_1$ when 
$p d <z < pd+d_t$ or $p d + d_t+d_2 < z < (p+1) d$ and $n(z)  = n_2$ 
when $p d+d_t <z < pd + d_t+d_2$. We choose $n_1 > n_2 > n_0$. 

For monochromatic TE waves the electric field is parallel to the OY 
axis. As in Ref. \cite{G1}, we write 
 \begin{equation}
{\bf E} = E_y(z) {\bf {\hat e}_y} e^{\displaystyle{i(k \beta x-\omega t)}}, 
\label{spie1}
\end{equation}
where $\omega$ is the angular frequency, $k=\omega/c$.
is the vacuum 
wavenumber and $\beta k$ is the (constant) $x$-component of the 
wavevector of modulus $k(z)=\omega n(z)/c$. 
For an electromagnetic 
wave entering the 1D photonic crystal from a uniform medium, one has 
\begin{equation}
\beta = n_0 \sin \theta_0,
\label{spie2}
\end{equation}
where $\theta_0$ is the angle of incidence measured from the normal. 
For monochromatic TM waves it is the magnetic field which lies parallel 
to the OY axis. Following Ref. \cite{G2}, we write 
 \begin{equation}
{\bf H} = H_y(z) {\bf {\hat e}_y} e^{\displaystyle{i(k \beta x-\omega t)}}.
\label{spie3}
\end{equation}
For piecewise constant $\epsilon(z) = n^2(z)$, 
the solutions of Maxwell's equations for $E(z)$ and $H(z)$ in the $p$-th 
cell are
\begin{eqnarray}
E_y(z),\,H_y(z) & = & a_p e^{ik_1(z-pd)} + b_p e^{-ik_1(z-pd)} \cr
& = & c_p e^{ik_2(z-pd)} + d_p e^{-ik_2(z-pd)}\cr
& = & e_p e^{ik_1(z-pd)} + f_p e^{-ik_1(z-pd)},
\label{spie4}
\end{eqnarray}
in regions $ pd < z< pd+d_t$, $ pd +d_t < z< pd+d_t+ d_2 $ and 
$pd+d_t+d_2 < z<  (p+1)d $ respectively. 
Of course, the coefficients $(a_p,\, b_p)$, $(c_p,\, d_p)$, and 
$(e_p,\, f_p)$ differ for TE and TM waves.

The transfer matrix, ${\cal M}$, is defined so that 
 \begin{equation}
\pmatrix{a_{p+1} \cr b_{p+1} } = {\cal M} \pmatrix{a_p \cr b_p}  
\equiv \pmatrix{A & B \cr B^* & A^*} \ \pmatrix{a_p \cr b_p}.
\label{spie5}
\end{equation}
Matching the fields in eq. (\ref{spie4}) and their derivatives 
(the derivatives divided by $n^2$) for TE (TM) waves, one finds
 \begin{eqnarray}
A^{\rm TE} & = & e^{ik_1 d_1} \left( \cos k_2 d_2 + {i\over 2} \left({k_1\over 
k_2} + {k_2 \over k_1} \right) \sin k_2 d_2 \right), \cr 
B^{\rm TE} & = & {i\over 2} e^{i k_1(d_1 -2 d_t) } \left({k_2 \over k_1}-
{k_1\over k_2}\right) \sin k_2 d_2,  
\label{spie6}
\end{eqnarray} 
and
 \begin{eqnarray}
A^{\rm TM} & = & e^{ik_1d_1} \left[ \cos k_2 d_2 + {i\over 2}
\left({{n_1^2k_2}\over {n_2^2 k_1}} + {{n_2^2 k_1}\over {n_1^2 
k_2}}\right) \sin k_2 d_2 \right],\cr 
B^{\rm TM} & = & {i\over 2} e^{ik_1(d_1-2 d_t)} \left({{n_1^2 
k_2}\over {n_2^2 k_1}}- {{n_2^2 k_1}\over {n_1^2 k_2}} \right) \sin k_2 d_2.
\label{spie7}
\end{eqnarray}
Once these are known, the Bloch waves of the 1D crystal are 
determined by the eigenvalue equation 
 \begin{equation}
\pmatrix{a_{p+1} \cr b_{p+1}} = \lambda \pmatrix{a_p \cr b_p} = 
{\cal M}  \pmatrix{a_p \cr b_p}~,  
\label{spie8}
\end{equation} 
and therefore the $\lambda$ satisfy 
 \begin{equation}
(A - \lambda) ( A^*- \lambda) - |B|^2 = 0.
\label{spie9}
\end{equation}
Using that det $ {\cal M} =|A|^2- |B|^2 = 1$, one finds
 \begin{equation}
\lambda_{\pm} = {\rm Re}(A) \pm \sqrt{{\rm Re}(A)^2 - 1},
\label{spie10}
\end{equation}
and eigenvectors 
 \begin{equation}
\pmatrix{ a \cr b} \propto \pmatrix{ B \cr \lambda_{\pm} -A}~. 
\label{spie11}
\end{equation}
In the bandgaps $\lambda_{\pm}$  are real since $|{\rm Re} A| > 1$.
In contrast, in allowed bands $\lambda_{\pm} = e^{\pm i \phi}$ with $\phi$ 
real. The bandgap boundaries are at Re$A = \pm 1$. 
Furthermore, from eq. (\ref{spie10}) $\lambda_- \lambda_+ = 1$. 
For surface states one chooses the $\lambda$ that corresponds to a  
damped Bloch wave when $z \to + \infty$. It must fulfill the 
condition $|\lambda| < 1$, and is $\lambda_-$ ( $\lambda_+$ ) when 
Re$(A) > 1$ ( $< -1$.) For simplicity we write it simply as $\lambda$ 
from here on. 

\subsection{Brewster Points}
Brewster points exist only for TM waves, and require Re$\,A^{\rm TM} = \pm 1$. 
{}From eq. (\ref{spie7}) we find that 
 \begin{eqnarray}
{\rm Re}\, A^{\rm TM} & = & \cos k_1d_1 \cos k_2 d_2 \cr
&-&{1\over 2} 
\left({{n_1^2 k_2}\over{n_2^2k_1}} + {{n_2^2k_1 }\over{n_1^2 k_2}} 
\right) \sin k_1 d_1 \sin k_2 d_2. 
\label{spie12}
\end{eqnarray}
It can be easily checked that a Brewster point occurs when 
 \begin{equation}
{k_1\over n_1^2 }={k_2\over n_2^2} 
\qquad {\rm and} \qquad 
k_1 d_1 + k_2 d_2 = m \pi,
\label{spie13}
\end{equation}
with $m = 1, 2, ...$ an integer which we  assume to be the 
bandgap index. The first of these equations determines $\beta$.
In particular, we have
 \begin{equation}
{{\sqrt{n_1^2- \beta^2}} \over n_1^2} = {{\sqrt{n_2^2- \beta^2}} \over n_2^2} 
\label{spie14}
\end{equation}
and, as a result,
 \begin{equation}
\beta_{Br} = {{n_1 n_2}\over{\sqrt{n_1^2+ n_2^2}}}\,.
\label{spie15}
\end{equation}
The second of equations (\ref{spie13}) then fixes the value of $k$ to be 
 \begin{equation}
k_{Br} = {{m\pi}\over{ d_1 \sqrt{n_1^2 -\beta^2}+ d_2 \sqrt{n_2^2-
\beta^2}}}~.
\label{spie16}
\end{equation}
For our reference case, see Fig. \ref{dsfig01},  we find that $\beta_{Br} = 
1.2$, $k_{Br} = 0.00816 \ nm^{-1}$ when $m=1$ (first bandgap). 
A numerical determination of the bandgap boundaries confirms that 
the gap width shrinks to zero at this point.

\section{Surface TM waves}

Most derivations are analogous to TE case, see Ref. \cite{Martorell2006}.
The magnetic field of a surface TM wave is written as
 \begin{equation}
H_y(z) = a_s e^{ik_1 z} + b_s e^{-ik_1 z}, 
\label{spie17}
\end{equation}
when $-d_s < z <0$ and 
 \begin{equation}
H_y(z) = b_v e^{q_0 z}~, 
\label{spie18}
\end{equation}
with $q_0 = + k \sqrt{\beta^2 - n_0^2}$, when $z < -d_s$. Using the 
boundary conditions, we obtain the exact dispersion relation 
$k=k(\beta)$ for TM surface waves by solving 
\begin{eqnarray}
{q_0\over k_1} \ {n_1^2\over n_0^2}& = & -i  \ {{\lambda^{\rm TM} -A^{\rm TM}
-{\tilde B}^{\rm TM}}\over{\lambda^{\rm TM}  -A^{\rm TM} +{\tilde B}^{\rm TM}}}, \cr 
{\tilde B^{\rm TM}} & \equiv & e^{-2ik_1 d_s} B^{\rm TM}. 
\label{spie19}
\end{eqnarray}
This equation must be solved numerically, and we will refer to the 
solutions thereby obtained as ``exact''. We note that the ratio 
$q_0/k_1$ depends only on $\beta$, and is independent of $k$. 
Furthermore, from eq. (\ref{spie9}) one has that $\lambda -A$, $B$ 
and ${\tilde B}$, have the same modulus and therefore eq. 
(\ref{spie19}) becomes: 
 \begin{equation}
{q_0\over k_1} \ {n_1^2\over n_0^2} = -i {{e^{i \theta_{\lambda-A} } 
- e^{i \theta_{\tilde B}}}  \over {e^{i \theta_{\lambda-A} } + e^{i 
\theta_{\tilde B}}}  } = \tan \left({{\theta_{\lambda-A} - 
\theta_{\tilde B}}\over 2}\right)~,
\label{spie20}
\end{equation}
where we have defined 
 \begin{equation}
\theta_{\lambda-A} \equiv {\rm arg}(\lambda -A) \qquad , \qquad 
\theta_{\tilde B} = {\rm arg} ({\tilde B}).
\label{spie21}
\end{equation}
To determine the latter, we note that 
 \begin{eqnarray}
c_- & \equiv & {{n_1^2 k_2}\over{n_2^2 k_1}} - {{n_2^2 k_1}\over{n_1^2 k_2}} \cr
& = & {{(n_1^2- n_2^2) (n_1^2 n_2^2- \beta^2(n_1^2+n_2^2))} \over 
{n_1^2n_2^2 \sqrt{(n_1^2-\beta^2)(n_2^2-\beta^2)}}}, 
\label{spie22}
\end{eqnarray}
appearing in eq. (\ref{spie7}) can have either sign. In fact we expect 
that the sign will change when $\beta$ runs from $n_0=1$ to $n_2$. 
We therefore write 
 \begin{equation}
\theta_{\tilde B} = {\pi \over 2} + k n_{1\beta} (d_1 -2 d_c) + \phi_s
\label{spie23}
\end{equation}
with $\phi_s$ chosen to be $0$ or $-\pi$ depending on the sign of 
$c_- \sin (k_2 d_2)$. As shown in Ref. \cite{Gaspar2004}, this sign is 
characteristic of each bandgap, unless the width shrinks to zero 
in either an optical hole \cite{Gaspar2004} or a Brewster point.
Returning to eq. (\ref{spie20}), we can now write $\theta_{\lambda-
A}$ explicitly as 
 \begin{equation}
\theta_{\lambda-A}(k) =  2  \Theta(\beta)+ k n_{1\beta} (d_1 -2 d_c) 
+ \phi_s + {\pi\over 2} + 2 \pi \nu 
\label{spie24}
\end{equation} 
with $  \nu = 0, \pm 1, \pm 2, ... \, $, and where we have defined 
 \begin{equation}
\Theta(\beta) \equiv \tan^{-1} \left( {n_1^2\over n_0^2} \sqrt{ 
{\beta^2 -n_0^2}\over{n_1^2 -\beta^2}}\right). 
\label{spie25}
\end{equation}
Note that all dependence on $k$ on the r.h.s. of eq. (\ref{spie24}) 
is explicit in the second term. Note also that in taking the inverse 
tangent we have introduced a contribution of $+ 2 \pi \nu$ on the 
r.h.s. This term was discussed at length in 
Ref.~\cite{Martorell2006}, so here we will discuss only the 
solutions with $\nu=0$. 

\begin{figure}[hbt]
\begin{center}
\begin{tabular}{c}
\includegraphics[height=7.5cm,width=8.5cm]{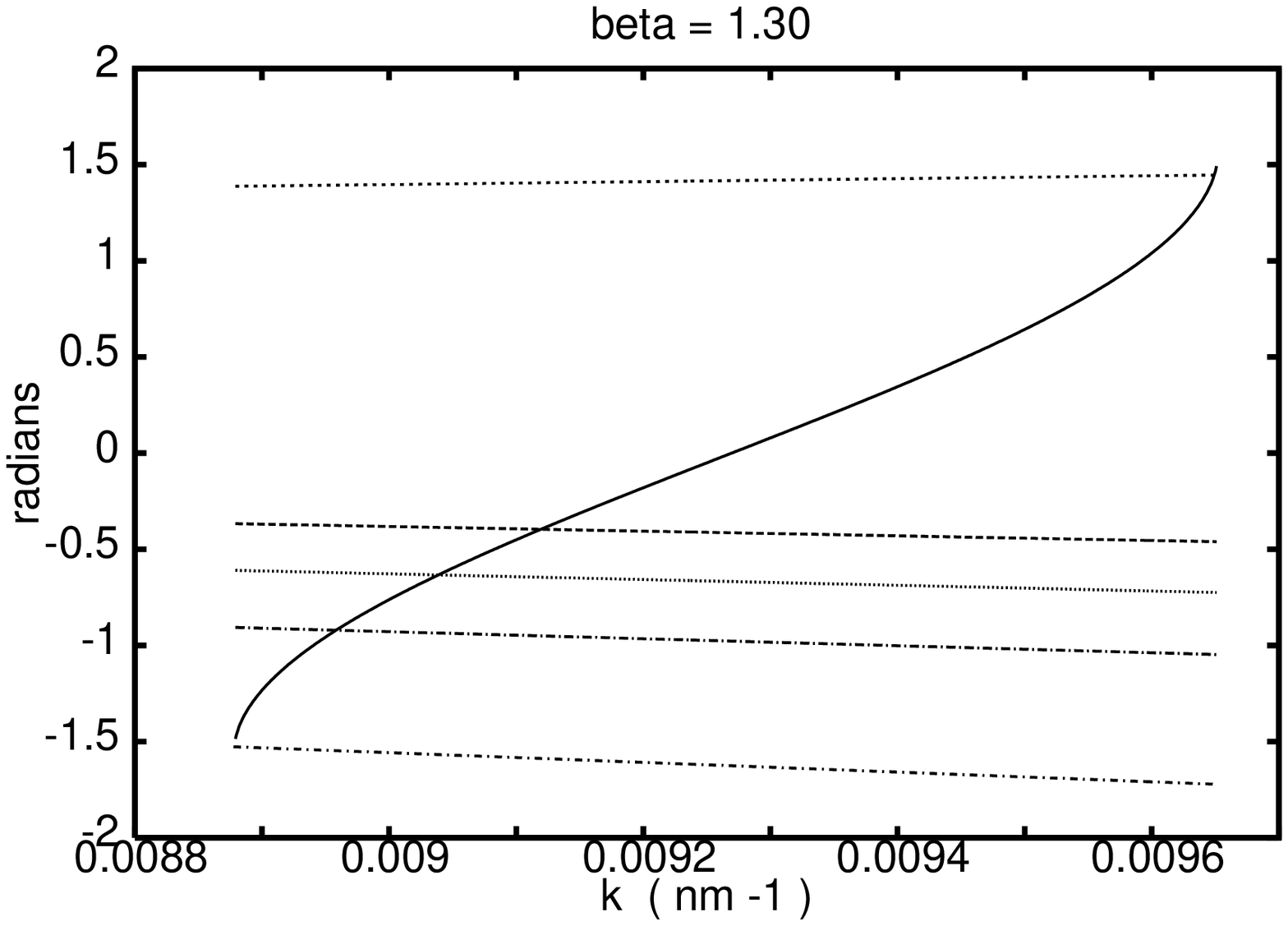}
\end{tabular}
\end{center}
\caption 
{\label{fi2tm130} 
Graphical solution of eq. \protect{(\ref{spie24})} for $n_0 = 1$, $n_1 
= 2$, $d_1 = 100 $ nm, $n_2 = 1.5$, $ d_2 = 250 $ nm, and $\beta = 
1.3$.  Continuous line: $\theta_{\lambda -A}$; dashed lines: 
r.h.s. for several values, from top to bottom, $d_c = 
0.25, 0.90, 0.99, 1.01$ and $1.33 d_1$.}   
\end{figure}

There is no simple analytic form for $\theta_{\lambda- A}$ as a 
function of $k$, so numerical methods must be used to solve eq. 
(\ref{spie24}). However, the results can be better understood using a 
simple graphical approach. In Fig. \ref{fi2tm130} we plot $ 
\theta_{\lambda-A}(k)$ (continuous line), and the r.h.s. of eq. 
(\ref{spie24}) for a chosen set of $d_c$'s (dashed lines). One sees 
that $\theta_{\lambda-A}$ increases from $-\pi/2$ to $\pi/2$ as $k$ 
ranges from the lower to the upper first bandgap boundaries. In a 
more general context, this property has been shown to hold in Ref. 
\cite{Sprung2004}. For a given $d_c$, the intersection of the 
corresponding straight line with the continuous line determines the 
solution for $k$. As the graph shows, when $d_c$ decreases, $k$ 
increases. The values of $d_c$ for which a solution can be found will 
therefore be bounded by those corresponding to the r.h.s. of eq. 
(\ref{spie24}) where it crosses $\theta_{\lambda-A} = \pm \pi/2$. 
That condition leads to 
\begin{eqnarray}
d_{c,min} & = & {d_1\over 2} + {1\over {k_R n_{1,\beta}}} 
\left( \Theta(\beta) + {\phi_s\over 2} \right)\,,\cr 
d_{c,max} & = & {d_1\over 2} + {1\over {k_L n_{1,\beta}}} 
\left( \Theta(\beta) + {\phi_s\over 2} + {\pi\over 2} \right). 
\label{spie26}
\end{eqnarray}
In the example shown in Fig. \ref{fi2tm130}, 
one finds $d_{c,min} = 21 $ nm and $d_{c,max} = 134 $ nm
(For this example, $\phi_s= -\pi$ and $\beta > \beta_{Br}$). 

\begin{figure}[hbt]
\begin{center}
\begin{tabular}{c}
\includegraphics[height=7.5cm,width=8.5cm]{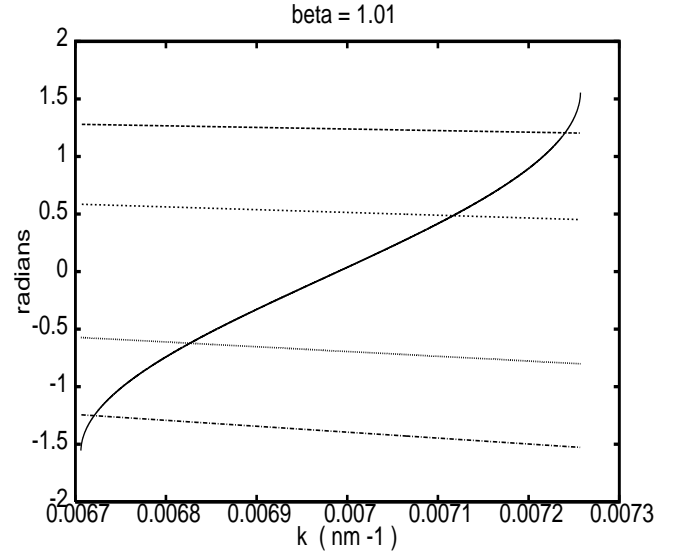}
\end{tabular}
\end{center}
\caption 
{\label{fi2tm101} 
Graphical solution of eq. \protect{(\ref{spie24})} for $n_0 = 1$, $n_1 
= 2$, $d_1 = 100 $ nm, $n_2 = 1.5$, $ d_2 = 250 $ nm, and $\beta = 
1.01$.   Continuous line: $\theta_{\lambda -A}$; dashed 
lines: r.h.s. for several values, from top to bottom, 
$d_c = 0.90, 1.20, 1.70$ and $2.0d_1$.} 
\end{figure}

The same expressions also apply 
when $\beta < \beta_{Br}$, but then $\phi_s =0$. If we choose 
$\beta= 1.01$, the limits are 
$d_{c,min} = 75 $ nm and $d_{c,max} = 134 $ nm. The graphical 
solutions are now shown in Fig. \ref{fi2tm101}. 

To proceed further in the analysis of the SEW solutions requires 
values for the band edges $k_L,\, k_R$ for a given $\beta$. These can 
be obtained from our semiclassical approximation \cite{G2}. In Ref. 
\cite{Martorell2006} we found very accurate analytic 
approximations for the TE bandgaps. Formally analogous expressions 
hold for TM bandgap boundaries, and their accuracy is 
also excellent. We have also shown that the semiclassical theory 
allows one to derive a good approximate expression for the argument of 
$\lambda -A$. In the first bandgap, it is 
 \begin{equation}
\theta_{\lambda-A}^{(e)} = \sin^{-1}\left({{k-k_m}\over {w/2}}\right) 
\label{spie27}
\end{equation}
with $ k_m \equiv (k_R + k_L)/2$ and $w = k_R-k_L$.  

\begin{figure}[hbt]
\begin{center}
\begin{tabular}{c}
\includegraphics[height=7.5cm,width=8.5cm]{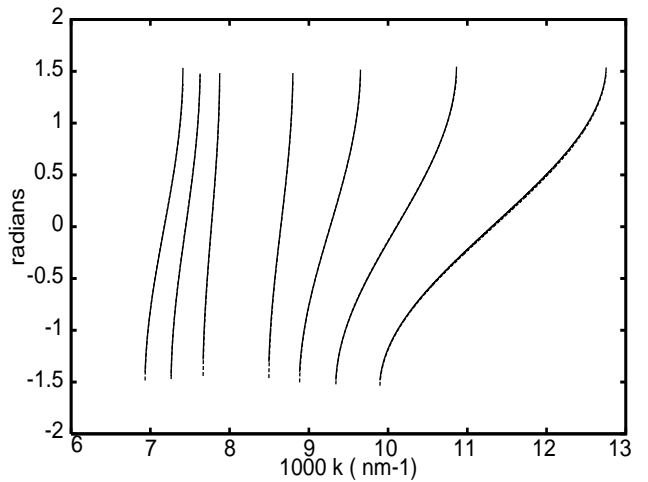}
\end{tabular}
\end{center}
\caption[example] 
{\label{tm4} 
Exact (continuous line) v.s. empirical approximation 
(dashed line) to the argument of $\lambda -A$. From left to right, 
$\beta = 1.05, 1.1, 1.15, 1.25, 1.3 , 1.35$ and $1.4$. We 
omitted the curves corresponding to the Brewster point at 
$\beta_{Br}= 1.2$ }  
\end{figure}

In Fig. \ref{tm4} we compare the exact valand the above approximation for 
$\theta_{\lambda-A}$ for several values of $\beta$ ranging from $n_0$ 
to $n_2$. For this example, the approximation is so good that on that scale one cannot 
distinguish between the exact and the approximate curves. 

\subsection{Approximate analytical solutions in the middle of the 
first bandgap.} 

\begin{figure}[hbt]
\begin{center}
\begin{tabular}{c}
\includegraphics[height=7cm,width=8.5cm]{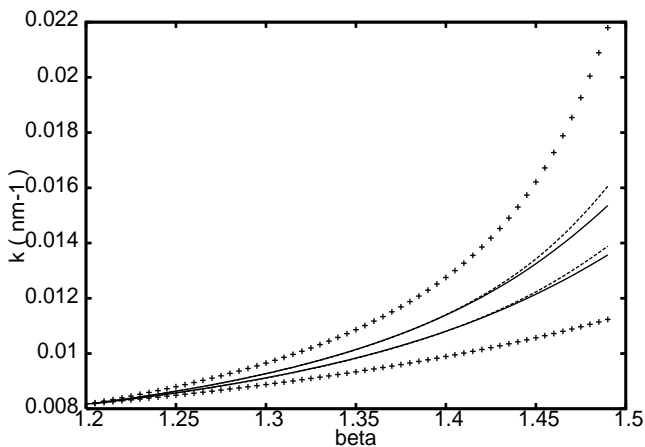}
\end{tabular}
\end{center}
\caption
{\label{tm5} 
Dispersion relation $k = k(\beta)$ when $d_c = 0.75\,d_1$, $0.90\,d_1$. 
Continuous line: exact solution of eq. \protect{(\ref{spie19})}. 
Dotted line: linear approximation, eq. \protect{(\ref{spie28})}.  
Crosses:  first bandgap boundaries. } 
\end{figure}

SEW in the middle of a bandgap are of  interest because the damping is strong, 
so most of the surface wave is confined very close to the 
surface.  For momenta such that  $k-k_m << w$ one can make the 
simplification $\sin^{-1} (2(k-k_m)/w) \simeq 2(k-k_m)/w$ and eq. 
(\ref{spie24}) then gives 
 \begin{equation}
k \simeq {{ \Theta(\beta) +   k_m/w + \phi_s/2 + \pi (\nu +1/4)}\over 
{1/w + n_{1\beta}( d_c -d_1/2 )}}~. 
\label{spie28}
\end{equation}
The dependence of $k$ on $d_c$, $\beta$ and $d_1$ is now transparent. 
The role of the bandgap parameters $k_m$ and $w$ can also be easily discussed.
Fig. \ref{tm5} shows the accuracy of this approximation 
when $\beta > \beta_{Br}$.

\subsection {Approximate solutions near the bandgap boundaries.}

\begin{figure}
\begin{center}
\begin{tabular}{c}
\includegraphics[height=7cm,width=8.5cm]{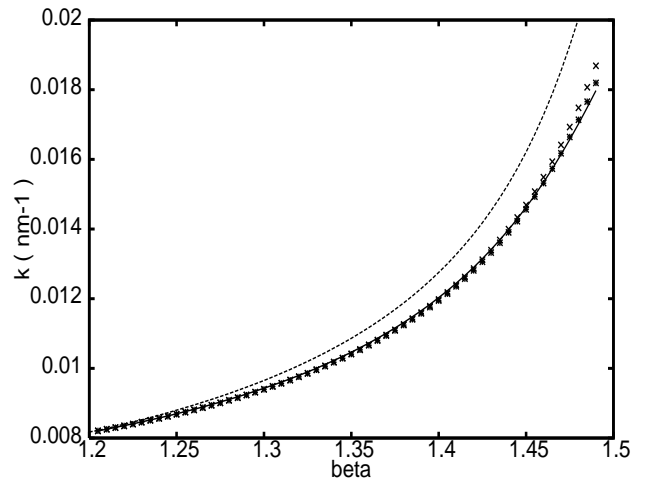}
\end{tabular}
\end{center}
\caption[example] 
{\label{tm6} 
Dispersion relation $k = k(\beta)$ when $d_c= 0.60\,d_1$ . 
Exact solution (continuous line) v.s. approximations :
Crosses: eq. \protect{(\ref{spie33})}; stars: eq. \protect{(\ref{spie34})}. 
The dashed line is the upper boundary of the first bandgap.} 
\end{figure}

Here we discuss solutions near the upper bandgap boundary, 
but similar approximations can be developed for the lower boundary, 
as in Ref. \cite{Martorell2006}. 
When $k$ is slightly below $k_R = k_m+ w/2$, it is convenient to 
introduce $\zeta > 0$ via  
 \begin{equation}
k- k_m = {w\over 2}(1-\zeta). 
\label{spie29}
\end{equation}
Then, 
 \begin{equation}
\sin^{-1} \left( {{k-k_m}\over {w/2}} \right) \simeq  {\pi \over 2} - \sqrt{2\zeta} 
={\pi \over 2} - 2 \ \sqrt{{k_R-k}\over w}. 
\label{spie30}
\end{equation} 
Inserting this into eq. (\ref{spie24}) gives 
 \begin{equation}
2 \ \sqrt{{k_R-k}\over w} + p (k_R-k) = \Lambda 
\label{spie31}
\end{equation}
with
 \begin{equation}
p \equiv n_{1\beta} ( 2 d_c -d_1)\, 
\quad 
\Lambda \equiv -2 \Theta - \phi_s + k_R p.
\label{spie32}
\end{equation}
Solving for $k_R - k$, we obtain 
 \begin{equation}
k = k_R - {1\over { p^2}}\left( -{1\over \sqrt{ w}} + \sqrt{{1\over w} 
+  p \Lambda} \right)^2 \ . 
\label{spie33}
\end{equation}
which is the desired solution, $k = k(\beta)$, near the upper bandgap 
boundary. Fig. \ref{tm6} shows an example of the accuracy of this 
expression. Furthermore, when $p \Lambda $ is small 
compared to $1/w$ one can expand and find 
 \begin{equation}
k = k_R - {1\over 4} w \Lambda ^2  \ ,
\label{spie34}
\end{equation}
which again manifests the dependence of $k$ on $d_c$ and $w$, 
and allows one to construct $k=k(\beta)$ very easily. Fig. \ref{tm6}  
shows again the validity of this approximation. 

\newpage
\section{First and Second Semiclassical Approximations}

The transfer matrix relates the amplitudes 
for transmission $t_B$ and reflection $r_B$ of a single cell as 
 \begin{equation}
\pmatrix{t_B \cr 0} = {\cal M} \pmatrix{1 \cr r_B}.
\label{spie35}
\end{equation}
Comparing to eq. (\ref{spie5}), $A = 1/t^*_B$ and $B = - r^*_B/t^*_B$ . 
In the first approximation, the expressions for $r_B$ and $t_B$ 
are given in Refs. \cite{Stolyarov1993,G1} 
for TE waves and in Ref. \cite{G2} for TM waves.
In particular, we have 
\begin{eqnarray}
r_B^{(1)} & = & {{-s_q^* \sinh (\gamma_1 d) }\over{\gamma_1 \cosh(\gamma_1 d) 
-i \delta_q \sinh(\gamma_1 d)}}\,,\cr
t_B^{(1)} & = & {{(-)^q \gamma_1}\over{\gamma_1 \cosh (\gamma_1 d) - i \delta_q 
\sinh(\gamma_1 d)}}\,,
\label{spie36}
\end{eqnarray}
where the coefficients $s_q$ differ of course for TE and TM waves.
For a photonic crystal with a symmetric cell, $d_1=2d_t$, they take the form
\begin{equation}
s^{\rm TE}_q  =   -\frac{i}{d}\ln\left[\frac{k_1}{k_2}\right] e^{-iq \pi}
\sin\left[\frac{d_2}{d}\left[q \pi + (k_2-k_1)d_1\right]\right]\,,
\label{spie37}
\end{equation}
\begin{equation}
s^{\rm TM}_q  =   -\frac{i}{d}\ln\left[\frac{k_1 \,n_2^2}{k_2 \,n_1^2}\right] e^{-iq \pi}
\sin\left[\frac{d_2}{d}\left[q \pi + (k_2-k_1)d_1\right]\right].
\label{spie38}
\end{equation}
The parameter $\delta_q$ is the detuning 
from the $q$-th Bragg resonance $k_q$, $q =1, 2...$ and 
the well-known Bragg condition for constructive interference is 
 \begin{equation}
k_q =\frac{\pi}{n_{av}d}\,\,q\,,\qquad \delta_q = kn_{av}-
\frac{\pi}{d}\,\,q~.
\label{spie39}
\end{equation}
The parameter $\gamma_1$ appearing in eqs. (\ref{spie36}) 
is defined as $\gamma_1 \equiv \sqrt{|s_q|^2- \delta_q^2}$\,.
Then, for the elements of the transfer matrix 
in the first approximation of the semiclassical coupled wave theory, 
we obtain
\begin{equation}
A^{(1)} = (-)^q \left( \cosh \gamma_1 d+ i{\delta_q \over \gamma_1} 
\sinh \gamma_1 d\right),
\label{spie40}
\end{equation}
\begin{equation}
B^{(1)} = (-)^q {s_q \over \gamma_1 } \sinh \gamma_1 d\,.
\label{spie41}
\end{equation}
Inserting  Re ($A^{(1)}$) into eq. (\ref{spie11}),  
we find that $\lambda^{(1)} = (-)^q e^{-\gamma_1 d}$,
where $\lambda^{(1)}$ is the first semiclassical approximation 
to the eigenvalue of the damped Bloch wave. \newpage

The second approximation introduced in Refs. \cite{G1,G2} 
leads to similar expressions for the elements of the transfer matrix:
\begin{eqnarray} 
A^{(2)} & = & (-)^q \bigg[ \cosh \gamma_2 d \cr
& + & i {{\left[(1+|u|^2)\eta_q - 2 {\rm Im}(s_q u^*)\right] }\over{(1-|u|^2 )\gamma_2}} \ \sinh \gamma_2 d \bigg],
\label{spie42} 
\end{eqnarray}
\begin{equation}
B^{(2)}= (-)^q \ {{s_q -2 i \eta_q u -s_q^* u^2}\over{(1-|u|^2)\gamma_2 }} \sinh \gamma_2 d\,, 
\label{spie43} 
\end{equation} 
where $\gamma_2 \equiv \sqrt{|s_q|^2- \eta_q^2}$,  $\eta_q =\delta_q - i c_2$ and 
\begin{equation}
c_2 = {{id}\over{2\pi}} \sum_{m \ne q= -
\infty}^{m=+\infty}{{|s_m|^2}\over{m-q-\delta_q d /\pi}}\,, 
\label{spie44}
\end{equation}
\begin{equation}
u =-\frac{id}{2\pi} \sum\limits_{m=-\infty,\, m\neq 
q}^{m=+\infty}\frac{s_m}{m-q-\delta_q\,d/\pi}\,.  
\label{spie45}
\end{equation}
The eigenvalue of the damped wave is $\lambda^{(2)} = (-)^q e^{-\gamma_2 d}$.

\begin{figure}[hbt]
\begin{center}
\begin{tabular}{c}
\includegraphics[height=7cm,width=8.5cm]{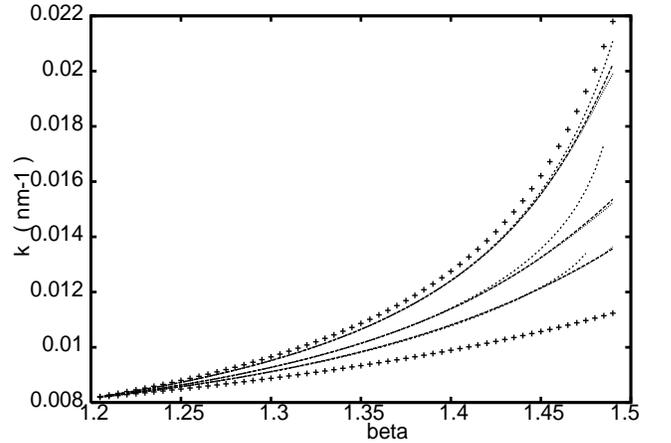}
\end{tabular}
\end{center}
\caption 
{\label{tm7} 
$k = k(\beta)$ curves for TM waves in the first bandgap 
for $d_c = 50, 75$ and $90$ nm; 
continuous lines: exact solutions; 
long dashes: first semiclassical approximations; 
short dashes: second semiclassical approximation; 
crosses: bandgap boundaries}
\end{figure}

Inserting $A^{(1,2)}$ and $B^{(1,2)}$ into eqs. (\ref{spie19}), one 
finds the corresponding predictions for $k=k(\beta)$ of TM surface 
waves. Selected results are shown in Fig. \ref{tm7}.  As can be seen, 
the second approximation gives excellent agreement with the exact 
dispersion relation within the range of validity $\beta < n_2$ of the 
semiclassical coupled wave theory, for the problem at hand. 

\begin{figure}
\begin{center}
\begin{tabular}{c}
\includegraphics[height=7cm,width=8.5cm]{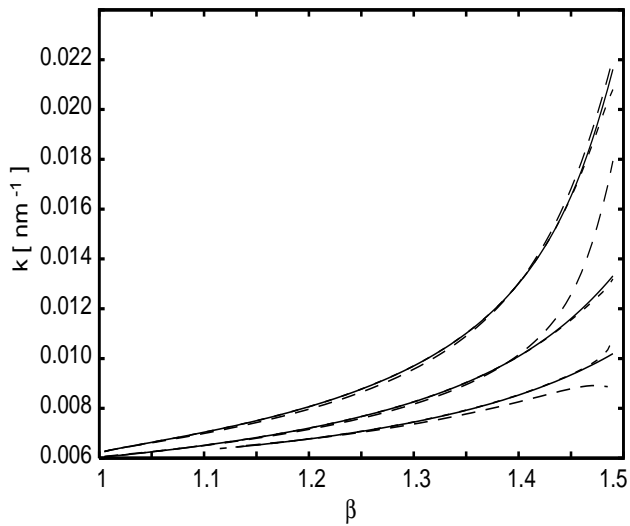}
\end{tabular}
\end{center}
\caption 
{\label{te8} 
$k = k(\beta)$ curves for TE waves in the first bandgap 
for $d_c = 25, 50 $ and $75$ nm; 
continuous lines: exact solutions; 
long dashes: first semiclassical approximations; 
short dashes: second semiclassical approximation. 
The latter curves are so close to the exact ones 
that the difference can be seen only when $\beta > 1.45$. }
\end{figure}

As shown in Ref. \cite{Martorell2006}, the dispersion relation 
$k=k(\beta)$ for TE surface waves is found by solving 
\begin{eqnarray}
{q_0\over k_1} & = & -i  \ {{\lambda^{\rm TE} -A^{\rm TE}
-{\tilde B}^{\rm TE}}\over{\lambda^{\rm TM}  -A^{\rm TE} +{\tilde B}^{\rm TE}}}, \cr 
{\tilde B^{\rm TE}} & \equiv & e^{-2ik_1 d_s} B^{\rm TE}. 
\label{spie46}
\end{eqnarray}
In Fig. \ref{te8} we compare the exact and semiclassical results 
for $k=k(\beta)$, choosing three thicknesses of the cap layer 
$d_c=25, 50$ and $75$ nm. The first approximation becomes 
inaccurate when $\beta$ exceeds $1.4$, but gives accurate results up 
to that value. The second approximation is so close to the exact 
values that one can see the difference only for values of $\beta$ 
very close to the critical value $\beta = n_2 = 1.5$. Beyond that, 
our semiclassical approximation cannot be applied, since $k_2$ 
becomes imaginary.

\acknowledgments       
We are grateful to NSERC-Canada for Discovery Grant RGPIN-3198 
(DWLS), and to DGES-Spain for continued support through grants  
BFM2001-3710 and FIS2004-03156 (JM). 


\end{document}